\documentclass[a4paper,fleqn]{cas-sc}

\usepackage[numbers]{natbib}

\def\tsc#1{\csdef{#1}{\textsc{\lowercase{#1}}\xspace}}
\tsc{WGM}
\tsc{QE}

\begin{document}
\let\WriteBookmarks\relax
\def\floatpagepagefraction{1}
\def\textpagefraction{.001}

\shorttitle{Potential Landscape Model of Interpersonal Communication}   

\shortauthors{G. N. A. Rahiemy}

\title [mode = title]{Breaking Down a Relationship Break-up: Potential Landscape as a Physics-Inspired Dynamical Model of Interpersonal Communication}  

\tnotemark[1] 


%

\author[1]{G. N. A. Rahiemy}[type=author,
                        orcid=0009-0005-5352-1197]

\cormark[1]

\ead{ghitha.rahiemy@tu-dortmund.de}



\affiliation[1]{organization={Department of Physics, TU Dortmund University},
            addressline={Otto-Hahn-Straße 4}, 
            city={Dortmund},
            postcode={44227}, 
            country={Germany}}

\cortext[1]{Corresponding author}



\begin{abstract}
A physics-inspired dynamical model of interpersonal communication is proposed to describe how information interpretation can influence the evolution of a relationship between two interacting systems. The model distinguishes between intended information and observed information, with the former represented as an asymptotic limit in an abstract information space. An information potential landscape with local minima and increasing potential barriers is introduced to represent stable information states and limited information accessibility. Information transmission and processing are described through a general thought-processing function, while the separation between information states determines the coupling strength between the interacting systems. The relationship dynamics is then defined from the coupling strength and the independent sentiments of the two systems. Numerical simulations demonstrate the potential landscape, information dynamics, and several relationship regimes under deterministic and fluctuating information processing. An illustrative dynamical scenario further shows that an emerging information discrepancy can lead to either relationship recovery or deterioration depending on the subsequent evolution of information separation, coupling, and individual sentiment. The model provides a simple physics-inspired framework for describing interpersonal communication as a coupled dynamical process.
\end{abstract}


\begin{highlights}
\item A potential landscape models stable interpersonal information states.
\item Information discrepancy modulates relational coupling.
\item Simulations illustrate relationship recovery and deterioration.
\end{highlights}


\maketitle

\section{Introduction}

People often say that humans are social beings. As social creatures, we naturally interact and communicate with one another throughout our lives. Communication is a broad activity through which individuals exchange signals, information, and meaning \cite{cherry1957human,ono2021interpersonal}. Although communication is usually discussed in the context of human interaction, interaction itself is not unique to humans. Physical systems continuously interact with their surroundings, exchange quantities, and change their states as a consequence of these interactions \cite{haas2015modeling}. This raises a simple question: can interpersonal communication also be understood as an interacting dynamical system?

Humans communicate through many different forms. Some are verbal, such as spoken or written language, while others are non-verbal, such as facial expressions, gestures, eye contact, physical touch, vocal characteristics, or even silence \cite{cherry1957human, ono2021interpersonal}. These observable forms of communication are produced by processes that are not directly visible to other people. Thoughts, memories, expectations, emotions, intentions, previous experiences, and physiological conditions may all influence what an individual eventually produces as an observable response \cite{cherry1957human, haas2015modeling, vasil2020world}.

This creates a fundamental limitation in interpersonal communication: an individual cannot directly observe another individual's complete internal state \cite{vasil2020world, palmer2015mental}. Instead, the individual receives observable outputs and must construct an interpretation from them \cite{cherry1957human, palmer2015mental}. In this sense, interpersonal communication is not simply the transfer of an internal state from one person to another. It is also an inference problem \cite{palmer2015mental, vasil2020world}.

This idea is central to several theories of human communication. Shannon's mathematical theory of communication provides a formal description of communication through a channel and explicitly considers the effects of noise and uncertainty during transmission \cite{Shannon1948}. However, interpersonal communication introduces an additional problem: the receiver does not merely decode a signal, but often has to infer what the sender intended by producing that signal \cite{Grice1975}. In this framework, the interpretation of an utterance depends not only on the signal itself but also on the cognitive context of the receiver. Therefore, the same observable output may produce different interpretations in different systems. This interpretation process has also been formalized using probabilistic models \cite{GoodmanFrank2016}. Such models provide an important conceptual basis for distinguishing between an original information state and the state reconstructed by another individual.

Several approaches have previously explored the modeling of communication, social interaction, cognition, and interpersonal dynamics using mathematical and physical frameworks. These include Shannon and Weaver's theory of communication \cite{Shannon1948}, statistical physics approaches to social dynamics \cite{castellano2009statistical, galesic2019statistical, bellingeri2026editorial, mulya2026destructive}, quantum models of cognition \cite{khrennikov2009quantum, lawless2017physics, busemeyer2025quantum, Khrennikov2020}, dynamical models and interpersonal information theory \cite{haas2015modeling, zhao2011entropy, perc2016phase}, models describing stability and oscillation in dyadic interaction \cite{felmlee1999dynamic}, and potential landscape approaches to social-ecological dynamics \cite{challis2026aggregate}. These studies demonstrate that mathematical and physics-inspired frameworks can capture different aspects of social interaction, cognition, and relational dynamics. The present work instead develops a minimal physics-inspired framework that combines intended and observed information, an abstract potential landscape for stable information states and accessibility barriers, and relationship dynamics based on information separation and individual sentiment. The aim is not to replace existing communication or social-dynamical theories, but to provide a simple framework for connecting information interpretation with interpersonal relationship dynamics.

The concept of local minima \cite{Coleman1977FalseVacuum1, Coleman1977FalseVacuum2, Berglund2013Kramers}, together with the concept of metastability in particle physics \cite{DenHollander2022Metastability, hiller2024vacuum}, provides a useful physical perspective for representing stable information states. A system can remain in a local minimum because it is surrounded by potential barriers, even when another state with lower potential exists. Based on this idea, an ``information potential landscape'' is introduced as an abstract space in which information can move during processing while stable information states are associated with local minima. Although the model does not aim to reproduce real human interaction in full detail, it provides a simplified representation of how information may evolve from its initial generation through transmission and interpretation, and how this interpretation may contribute to the resulting relationship state. Here, relationship break-up is regarded as a possible consequence of relationship deterioration, rather than as an explicitly defined state of the present model.

The proposed model has four main assumptions concerning how information is represented, where it is abstractly located, how it is shared, and how the relationship state changes over time. The model distinguishes between intended and observed information, with the latter representing an interpreted approximation of the former. Information processing is represented by a general thought-processing function that moves the information state through the landscape. The resulting information separation is then used to determine the relational coupling, which acts together with the independent sentiments of the two interacting systems to produce the relationship dynamics. Repeated interaction therefore generates an evolving trajectory through the information space, allowing different relationship states to emerge from the interplay between information discrepancy, coupling, and individual sentiment. 

\section{Model Building}

The overall conceptual structure of the proposed model can be summarized as information transmission,

\begin{equation}
\boxed{
\text{internal state}
\rightarrow
\text{observable output}
\rightarrow
\text{interpretation}
\rightarrow
\text{updated state}
\rightarrow
\text{new output}
}
\end{equation}

embedded within

\begin{equation}
\boxed{
\text{information landscape}
+
\text{information dynamics}
\to
\text{relationship dynamics}
}
\end{equation}

\subsection{Model Assumptions}

The assumptions were formulated based on concepts from psychology and communication science. The proposed model considers four main aspects:

\begin{enumerate}
    \item How information is represented: intended and observed information,
    \item Where information is abstractly located: the information potential landscape,
    \item How information is shared: information transmission,
    \item How the relationship state changes over time: relationship dynamics.
\end{enumerate}

\subsubsection{Intended and observed information}

In the present work, the problem of ``different perspectives in interpersonal interaction'' motivates the concept of \emph{intended information}. Intended information ($x_{\text{intended}}$) represents the information associated with the internal state that gives rise to an observable output. The information available to another individual is therefore not assumed to be identical to the original information state. The intended information could be placed at any given location within this information space. In the \textit{first} model, however, the intended information is represented as an asymptotic limit,

\begin{equation}
x_{\mathrm{intended}}\rightarrow\infty.
\end{equation}

For approximation, the information can instead be represented by a finite position. This gives rise to the concept of limited information accessibility. The underlying psychological motivation comes from the fact that interpretation depends on the receiver's cognitive state, context, and available knowledge \cite{SperberWilson1995, GoodmanFrank2016}. The model assumes that the information originally intended by the originating system is not perfectly accessible to the receiving system. This limitation is not necessarily restricted to communication between two different people. An individual may also have limited access to the internal processes that produce their own behaviour \cite{NisbettWilson1977}. This suggests that a distinction between an underlying state and an observable output may also exist within a single individual.

The asymptotic coordinate is an idealized representation of complete access to the originating information state. Since the receiving system can only access finite information states through observable communication, the observed information state remains an approximation of the intended information.

Observed information ($x_{\text{observed}}$) serves as the reconstructed representation formed by a receiving system based on the signals it receives. Therefore, within a one-dimensional space, the coordinate $x \ge 0$ represents the functional position of an information state in an abstract cognitive landscape. The origin, $x = 0$, corresponds to the baseline state of maximum misalignment, where the observed state is least aligned with the originating system's internal state. While the intended information is deliberately formulated as an asymptotic limit to reflect the cognitive impossibility of fully capturing or replicating another individual's unadulterated internal state, the observed information represents a finite approximation achieved through noisy communication channels and inferential thought processing.

\subsubsection{Information potential landscape}

To support the idea of an information potential landscape, information is assumed to remain in metastable states surrounded by potential barriers within an abstract potential landscape. Information is represented by a coordinate $x$ in a shared information space. For this \textit{first} model, this idea is implemented by describing the landscape as

\begin{equation}
V(x)=-e^{\alpha x}\cos{x},
\qquad
0<\alpha\leq1,
\qquad x\geq 0.
\label{eq:potential}
\end{equation}

The parameter $\alpha$ controls the rate at which the amplitude of the landscape increases with $x$. A larger $\alpha$ produces increasingly high barriers at larger values of $x$, representing decreasing accessibility to information states farther from the origin.

The functional form in Eq. \eqref{eq:potential} is selected as the simplest dynamical representation that captures both metastable storage and decreasing accessibility along the information coordinate:
\begin{enumerate}
    \item \textbf{Oscillatory Structure ($\cos x$):} The periodic spatial modulation generates a sequence of local minima (basins) separated by local maxima (potential barriers). The local minima represent stable, bound information states where a reconstructed interpretation can temporarily reside. The local maxima represent potential barrier between adjacent basins that acts as an effective energy threshold, classically defining the processing effort required for an information state to transition into a neighboring basin. Surrounded by these potential walls, the information is constrained from freely wandering across the landscape without sufficient forcing.
    \item \textbf{Exponential Envelope ($e^{\alpha x}$):} The exponential factor gradually amplifies the peak-to-valley amplitude along $x$. This ensures that as an information state attempts to traverse deeper into the landscape toward the asymptotic intended state $x_{\text{intended}}$, the potential barriers grow systematically higher. Consequently, accessing information states farther from the initial representation demands progressively higher cognitive processing energy, embodying the principle of limited information accessibility.
\end{enumerate}

An information state may move through the information space during processing, but stable information states are constrained to the set

\begin{equation}
\mathcal{I}
=
\left\{
x:
V'(x)=0,\,
V''(x)>0
\right\}.
\label{eq:manifold}
\end{equation}

Thus, the minima of $V(x)$ represent the positions at which information can remain stable after a processing step. The potential difference between a local minimum and the neighbouring maximum defines the transition barrier,

\begin{equation}
\Delta V_n
=
\left|
V_n^{\mathrm{max}}
-
V_n^{\mathrm{min}}
\right|.
\label{eq:barrier}
\end{equation}

The barrier represents the amount of effective processing required for an information state to transition from one stable basin to another. Under a classical transition interpretation, if

\begin{equation}
E_{\mathrm{proc}}<\Delta V_n,
\end{equation}

the transition to the next basin is inaccessible. Conversely, when

\begin{equation}
E_{\mathrm{proc}}\geq\Delta V_n,
\end{equation}

the transition becomes accessible.

The quantity $\Delta V_n$ therefore represents a model quantity describing the increasing difficulty of accessing information states farther from the initial representation.

\subsubsection{Information transmission}

The model then consists of two interacting systems, $S_1$ and $S_2$. Each system possesses an internal state and produces an observable output through a thought-processing function $F_i$. The output of one system becomes the interpreted input of the other system, where the transmitted information may already differ from the original intended information due to the effects of transmission, interpretation, and contextual conditions. In its simplest form, the interaction is written as

\begin{equation}
S_1
\rightarrow
I_{\mathrm{intended}}
\xrightarrow{F_1}
x_{\mathrm{intended}}^1
\xrightarrow{\mathcal{T}}
x_{\mathrm{observed}}^1
\rightarrow
S_2
\xrightarrow{F_2}
x_{\mathrm{inferred}}^2
\xrightarrow{\mathcal{T}}
x_{\mathrm{observed}}^2
\rightarrow
S_1.
\end{equation}

where $\mathcal{T}$ represents the combined effects of information transmission, interpretation, previously available information, and environmental or contextual conditions. The function $F_i$ is intentionally kept general in the initial model. It represents the process through which an individual receives information, combines it with information already available within the system, and produces a new output. It is also possible for the processing function to generate an already modified output from the initial internal state, based on the assumption that an individual may have limited access to the internal processes that produce their own behaviour \cite{NisbettWilson1977}.

The interpersonal interaction begins when a system with state $S_1$ produces an intended information $I_{\mathrm{intended}}$, which is processed by the internal function $F_1$ and mapped from the internal system into the shared information space. The resulting information occupies a certain position $x_{\mathrm{intended}}^1$ in the information landscape. The second system, $S_2$, receives this information through the transmission process $\mathcal{T}$ and obtains an observed information state that can differ from the intended information. In the simplest representation, this transformation can be written as

\begin{equation}
x_{\mathrm{observed}}^1
=
x_{\mathrm{intended}}^1-\eta(x),
\label{eq:info_transformation}
\end{equation}

where $\eta(x) \neq 0$ represents the net effect of transmission and environmental conditions that shifts the position of the observed information. Therefore,

\begin{equation}
x_{\mathrm{intended}}^1
\neq
x_{\mathrm{observed}}^1.
\end{equation}
This information transmission between systems is motivated by the inferential nature of communication, collaborative grounding, and the dependence of interpretation on context and previous knowledge \cite{Grice1975, ClarkBrennan1991, ClarkSchaefer1989, SperberWilson1995, PickeringGarrod2004}.

The observed information then enters the internal state of $S_2$ and is processed by $F_2$. The resulting processing changes the position of the information in the shared information space. In the initial model, this temporal evolution is described by

\begin{equation}
\frac{dx}{dt}
=
\Phi_i(t),
\label{eq:dynamics}
\end{equation}

where $\Phi_i(t)$ represents the effective information-processing forcing produced by system $i$. The effective forcing is related to the underlying thought-processing function through

\begin{equation}
\Phi_i(t)
=
F_i
\left[
S_i(t),
I_i^{\mathrm{past}}(t),
J_i(t)
\right],
\label{eq:forcing}
\end{equation}

where $S_i(t)$ represents the current internal state of system $i$, $I_i^{\mathrm{past}}(t)$ represents previously stored information, and $J_i(t)$ represents newly received information. Therefore, $\Phi_i(t)$ represents the net effect of these components on the current information trajectory. The formulation is intentionally general because the detailed structure of the thought-processing function is beyond the scope of the first model.

For a given initial condition $x_0$, the resulting trajectory can be written as

\begin{equation}
x^*(t)
=
x_0+\int_0^t \Phi_i(\tau)\,d\tau.
\label{eq:candidate}
\end{equation}

The trajectory $x^*(t)$ describes the temporal evolution of the information state during processing. It is not required to remain at a stable information state at every moment. However, A position is regarded as a stable information state only when it coincides with a local minimum of \(V(x)\). Intermediate positions therefore represent transient states during processing rather than stable information states.

Thus, the information dynamics describes the movement between states, whereas the potential landscape determines the stable positions at which information can eventually settle. Repeated interaction therefore produces a trajectory through the information landscape as the information is repeatedly processed and reinterpreted.

\subsubsection{Relationship Dynamics}

Repeated information exchange between coupled systems can generate distinct dynamical regimes depending on the information-processing dynamics and the coupling between the systems. This hypothesis is motivated by the observation that interpersonal interaction is iterative and state-dependent. The argument is supported by interactive alignment theory
\cite{PickeringGarrod2004}, Communication Accommodation Theory \cite{Giles1973,GilesEdwardsWalther2023}, and Relational Turbulence Theory \cite{SolomonKnoblochTheissMcLaren2016}, as well as related work on collaborative grounding and mutual understanding \cite{ClarkBrennan1991,ClarkSchaefer1989}, conceptual alignment in conversation \cite{BrennanClark1996}, uncertainty reduction and relationship development \cite{BergerCalabrese1975,AltmanTaylor1973}, collaborative reference and grounding \cite{ClarkBrennan1991,ClarkSchaefer1989,ClarkWilkesGibbs1986}, and conversational repair \cite{SchegloffJeffersonSacks1977}.

The relationship dynamics in the proposed model is defined as

\begin{equation}
R(t)
=
K_{12}(t)
\left[
S_1(t)+S_2(t)
\right],
\label{eq:relationship}
\end{equation}

where $K(t)_{12}$ is the relational coupling strength as a joint property of the two systems, while $S_1(t)$ and $S_2(t)$ respectively represent the individual and independent sentiments of system 1 and system 2 toward the relationship. The coupling is further described by

\begin{equation}
\left| K_{12}(t) \right|
=
K_0-\beta D_{IO}(t),
\label{eq:coupling}
\end{equation}

where $D_{IO}(t)$ is the information separation between the two systems, $K_0$ represents baseline factors that maintain the coupling, such as history, familiarity, attachment, shared experience, or other factors that strengthen the relationship, and $\beta$ represents the effect of information separation on the coupling. The relational coupling strength is taken to be non-negative in the present model, such that the dynamical direction of the relationship is determined by the combination of $S_1(t)$ and $S_2(t)$. The separation between the information states of the two systems is defined as

\begin{equation}
D_{IO}(t)
=
\left|
x_I-x_O(t)
\right|.
\label{eq:systemseparation}
\end{equation}

where $x_I$ is the intended information and $x_O(t)$ refers to the observed coordinate, both in the abstract information space $\mathcal{I}$ at a certain time $t$. A small value of $D_{IO}$ represents similar information representations, whereas a large value represents a greater difference between the information states of the two systems. In the present minimal formulation, information discrepancy modulates the magnitude of relational coupling, while the direction of relationship dynamics is determined by the combined individual sentiments.

Several relationship states can therefore emerge from the proposed model:

\begin{enumerate}
    \item \textbf{Absolute positive relation:} All components, $K_{12}$, $S_1$, and $S_2$, are positive, resulting in positive relationship dynamics.

    \item \textbf{Absolute negative relation:} Both $S_1$ and $S_2$ have negative sentiments. Since $K_{12}$ is non-negative, the magnitude of the coupling determines how strongly the negative sentiment is expressed in the relationship dynamics, resulting in a negative relationship state.

    \item \textbf{One-sided negative relation:} One of $S_1$ or $S_2$ is negative, while the other is positive, but the negative sentiment is sufficiently strong that
    \begin{equation}
    S_1+S_2<0.
    \end{equation}
    The relationship therefore remains in a negative state. In a real situation, this case could represent a condition in which the negative sentiment of one system is strong enough that the other system cannot neutralize it nor maintain a positive overall relationship state.

    \item \textbf{One-sided positive relation:} One of $S_1$ or $S_2$ is negative while the other is positive, but the positive sentiment is sufficiently strong that
    \begin{equation}
    S_1+S_2>0.
    \end{equation}
    The resulting relationship dynamics therefore remains positive.

    \item \textbf{Neutral relationship:} A neutral relationship can occur when
    \begin{equation}
    K_{12}=0.
    \end{equation}
    In this case, regardless of the individual sentiment states of the two systems, the relationship dynamics remains zero because there is no effective coupling between them. This can represent a situation in which no interaction or information sharing has occurred, and therefore no relational dynamics is generated within the model.
\end{enumerate}

\subsection{Numerical Implementation}

The numerical implementation is designed to provide a simple first realization of the proposed model while preserving the distinction between the information landscape, information dynamics, and relationship dynamics. The calculations are performed over a finite numerical domain, since the intended information is represented by the asymptotic limit $x_{\mathrm{intended}}\rightarrow\infty$ in the analytical model. A finite value of $x_{\max}$ is therefore used for numerical evaluation.

\subsubsection{Information Potential Landscape}
The potential landscape is evaluated over

\begin{equation}
0\leq x\leq x_{\max},
\end{equation}

for several values of the accessibility deterioration parameter $\alpha$. The extrema of the potential are obtained from

\begin{equation}
V'(x)=0,
\end{equation}

and classified using the sign of the second derivative,

\begin{equation}
V''(x)
\begin{cases}
>0, & \text{local minimum},\\
<0, & \text{local maximum}.
\end{cases}
\end{equation}

The corresponding potential difference between a local minimum and the neighbouring maximum is then used to characterize the transition barrier,

\begin{equation}
\Delta V_n
=
\left|
V_n^{\mathrm{max}}
-
V_n^{\mathrm{min}}
\right|.
\end{equation}

\subsubsection{Information Dynamics}

The information transformation in Eq. \eqref{eq:info_transformation} indicates that the condition $\eta(x) \neq 0$ applies strictly to the ideal asymptotic formulation. In contrast, for the finite numerical approximation, $\eta = 0$ is permitted when the observed state reaches the finite representation of the intended state.

For the information dynamics, the continuous equation

\begin{equation}
\dot{x}
=
\Phi(t)
\end{equation}

is discretized using a time step $\Delta t$. A first-order Euler update is used,

\begin{equation}
x_{k+1}
=
x_k+\Phi(t_k)\Delta t.
\label{eq:euler}
\end{equation}

To acquire the exact information position within the information potential landscape, we derive the potential formulation in Eq. \eqref{eq:potential} by considering its differential and applying the stable information states constrained in Eq. \eqref{eq:manifold}:

\begin{equation}
    x_k = \arctan(\alpha) + 2\pi k,
\end{equation}
where $k$ is an integer ranging from $0$ to $10$ in this simulation.

The simple numerical implementation considers three information-processing regimes:

\begin{enumerate}
    \item \textbf{No update:}
    \begin{equation}
    \Phi(t)=0,
    \label{eq:noup}
    \end{equation}
    corresponding to a state in which the information position does not change with time.

    \item \textbf{Constant update:}
    \begin{equation}
    \Phi(t)=A,
    \label{eq:constant}
    \end{equation}
    where $A$ is a constant information-processing rate. Positive $A$ produces an upward information trajectory, while negative $A$ produces a downward trajectory.

    \item \textbf{Time-dependent update:}
    \begin{equation}
    \Phi(t)=A(t),
    \label{eq:timedep}
    \end{equation}
    which on this modeling, we choose;
    \begin{equation}
        A(t) = A \sin(\omega t)
    \end{equation}
    with $A=0.1$ and $\omega=1.0$. Here, the information-processing rate changes with time and can therefore produce alternating increases and decreases in the information coordinate.
\end{enumerate}

The numerical information space is restricted to $x\geq0$. In addition, the intended information coordinate is treated as an absorbing upper boundary in the numerical approximation. Once the observed information, $x_O$, reaches the finite approximation of the intended coordinate, $x_I$, it remains at that position,

\begin{equation}
x_O(t)=x_I
\qquad
\text{for}
\qquad
x_O(t)\geq x_I.
\end{equation}

This implements the assumption that the system cannot move beyond the intended-information state in the initial model. Consequently, once

\begin{equation}
D_{IO}(t)
=
|x_I-x_O(t)|
=
0,
\end{equation}
the discrepancy remains zero for the remainder of the processing.

For the first relationship-dynamics simulations, the normalized information separation is calculated under three simplified information-processing conditions and therefore it only exists in the range $0 \leq D_{IO} \leq 1$. The third case uses a time-dependent uniform random information-processing rate generated within

\begin{equation}
-150\leq A(t)\leq150,
\end{equation}

to represent fluctuating information-processing dynamics. The symmetric range is intentionally selected so that the information state can move in both directions with comparable magnitudes.

\subsubsection{Relationship Dynamics}
The resulting information separation is then incorporated into the relational coupling as stated in Eq. \eqref{eq:coupling} and the relationship dynamics is evaluated for the five relationship states described above. This procedure allows the effect of information separation on the relationship dynamics to be examined while keeping the individual sentiment states controlled.

To further investigate temporal fluctuations in the relationship state, a stochastic realization is generated for each relationship regime. In the $j$-th realization, the relationship dynamics is calculated as

\begin{equation}
R^{(j)}(t)
=
K_{12}^{(j)}(t)
\left[
S_1^{(j)}(t)
+
S_2^{(j)}(t)
\right].
\label{eq:relationship_realization}
\end{equation}

Here, the information-processing trajectory and individual sentiment states are independently regenerated for each realization. The ensemble of realizations provides a statistical description of the relationship dynamics rather than a single deterministic trajectory.

The mean relationship dynamics is calculated over $N_{\mathrm{real}}$ realizations as

\begin{equation}
\langle R(t)\rangle
=
\frac{1}{N_{\mathrm{real}}}
\sum_{j=1}^{N_{\mathrm{real}}}
R^{(j)}(t),
\label{eq:mean_relationship}
\end{equation}

while the temporal standard deviation is calculated as

\begin{equation}
\sigma_R(t)
=
\sqrt{
\left\langle R(t)^2\right\rangle
-
\left\langle R(t)\right\rangle^2
}.
\label{eq:std_relationship}
\end{equation}

The mean represents the average dynamical trend of the relationship over the ensemble of realizations, while $\sigma_R(t)$ measures the spread of the relationship dynamics at each time. In the resulting plots, the mean trajectory is used to represent the central dynamical trend, while the interval $\langle R(t)\rangle\pm\sigma_R(t)$ represents the corresponding statistical fluctuation.

\begin{figure}[htbp]
    \centering
    \includegraphics[width=0.75\linewidth]{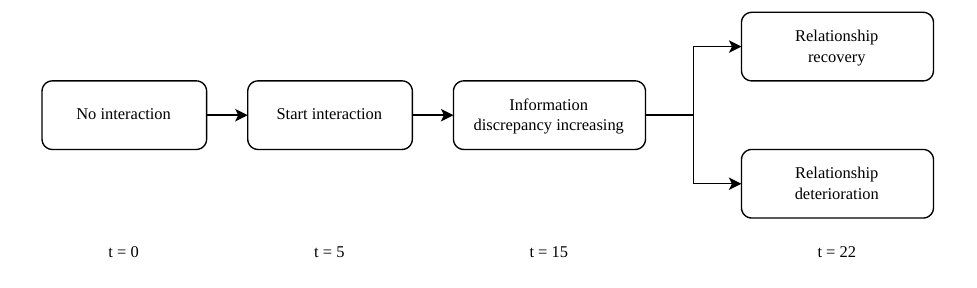}
    \caption{
    Flow chart of the illustrative relationship scenario, showing the
    transition from no interaction to interaction, the emergence of
    information discrepancy, and the subsequent branching into relationship
    recovery or deterioration.
    }
    \label{fig:scenario_flowchart}
\end{figure}

Finally, a more illustrative relationship trajectory is constructed to demonstrate how the system can evolve from a nearly neutral initial state into different relationship outcomes. The simulation begins with a small baseline coupling and weak information interaction. Random information processing then produces small fluctuations in $D_{IO}$. After an information discrepancy emerges, one individual sentiment is allowed to become negative, creating a state from which two qualitative branches can develop. In the recovery branch, information separation decreases, the baseline coupling $K_0$ increases, and the negative sentiment is gradually neutralized. In the deterioration branch, the information separation continues to increase, $K_0$ remains approximately unchanged, and both individual sentiments become increasingly negative. These two branches illustrate how the same initial interaction can lead to qualitatively different relationship dynamics depending on the subsequent evolution of information separation, coupling, and individual sentiment. The flow of this illustrative relationship scenario is shown in Fig.~\ref{fig:scenario_flowchart}.

The numerical simulations were performed using the parameters summarized in Table~\ref{tab:numerical_parameters}. The parameters are separated according to the three numerical experiments: information discrepancy, statistical relationship dynamics, and the illustrative relationship scenario. The information-discrepancy simulation uses the first and final local minima of the potential landscape as the initial and intended information positions, respectively. The relationship-dynamics simulations use independently generated information and sentiment fluctuations over multiple realizations.

\begin{table}[htbp]
    \centering
    \caption{
    Parameters used in the numerical simulations of information discrepancy,
    statistical relationship dynamics, and the illustrative relationship
    scenario.
    }
    \label{tab:numerical_parameters}
    \begin{tabular}{l l c}
        \hline
        \textbf{Simulation} & \textbf{Parameter} & \textbf{Value} \\
        \hline

        \multirow{5}{*}{Information discrepancy}
        & $\alpha$ & 0.1 \\
        & $\Delta t$ & $0.01$ \\
        & $A$ & $0.05$ \\

        \hline

        \multirow{7}{*}{Statistical relationship dynamics}
        & $T$ & $30$ \\
        & $N$ & $1000$ \\
        & $N_{\mathrm{real}}$ & $500$ \\
        & $x_I$ & $5.0$ \\
        & $x_O(t=0)$ & $0.0$ \\
        & $K_0$ & $1.0$ \\
        & $\beta$ & $0.8$ \\

        \hline

        \multirow{8}{*}{Illustrative scenario}
        & $T$ & $40$ \\
        & $N$ & $1200$ \\
        & $N_{\mathrm{real}}$ & $500$ \\
        & $x_I$ & $5.0$ \\
        & $\beta$ & $0.55$ \\
        & $K_{0,\mathrm{initial}}$ & $0.12$ \\
        & $t_{\mathrm{interaction}}$ & $5$ \\
        & $t_{\mathrm{discrepancy}}$ & $15$ \\
        & $t_{\mathrm{branch}}$ & $22$ \\

        \hline
    \end{tabular}
\end{table}
The notation used throughout the numerical implementation follows the definitions introduced in the model, where $\alpha$ denotes landscape amplitude rate, $t$ denotes time, $N$ the number of time points, $N_{\mathrm{real}}$ the number of realizations in statistical simulation, $x_I$ is the intended information positions, $x_O(t=0)$ the initial observed information position, $A$ the information-processing rate, $\Delta t$ the time step, $K_0$ the baseline coupling, and $\beta$ the information-separation coupling parameter.

\section{Results and Discussion}

\subsection{Information Potential Landscape}

The information potential landscape defined by Eq.~(\ref{eq:potential}) was first evaluated to investigate the structure of the proposed information space. The landscape was initially plotted over the interval $0\leq x\leq10$ for $\alpha=1$.

\begin{figure}[htbp]
    \centering
    \includegraphics[width=\linewidth]{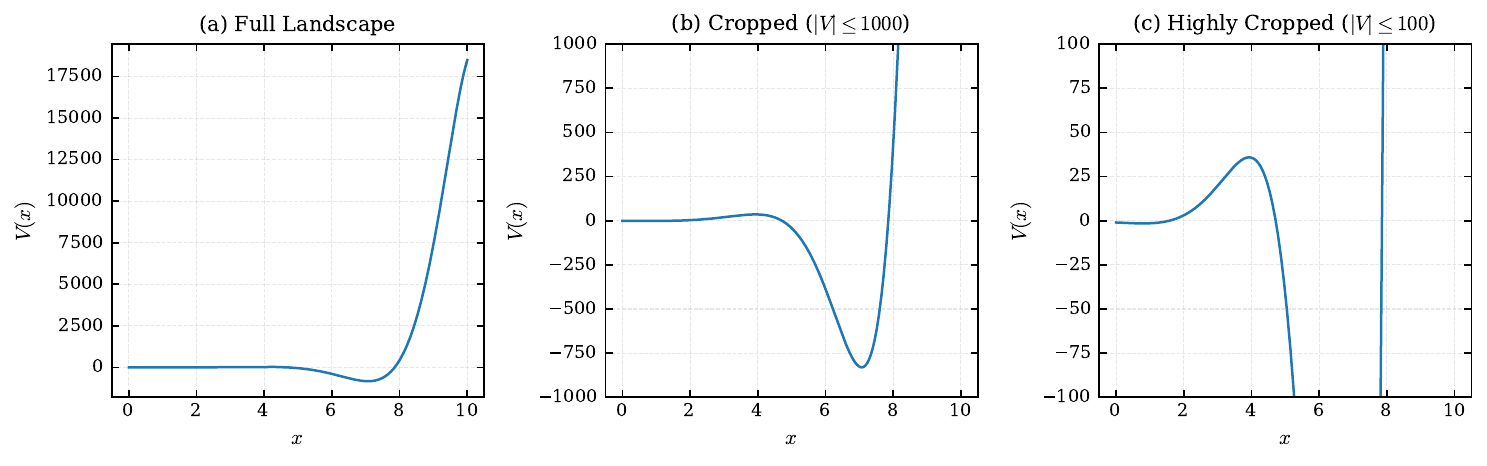}
    \caption{
    Information potential landscape for $\alpha=1$ over the interval
    $0\leq x\leq10$. The exponential envelope causes the magnitude of
    the potential to increase rapidly with $x$, making the later local
    minima and maxima difficult to distinguish visually without
    inspecting the corresponding numerical values.
    }
    \label{fig:landscape_panels}
\end{figure}

As shown in Fig.~\ref{fig:landscape_panels}, the oscillatory structure is already visible at small values of $x$. However, the exponential factor causes the magnitude of the potential to increase rapidly as $x$ increases. Consequently, later extrema become increasingly large in magnitude and the earlier minima and barriers become visually compressed when plotted over a wider range. The landscape was therefore extended to $0\leq x\leq30$ to examine the behaviour of the later extrema.

\begin{figure}[htbp]
    \centering
    \includegraphics[width=0.6\linewidth]{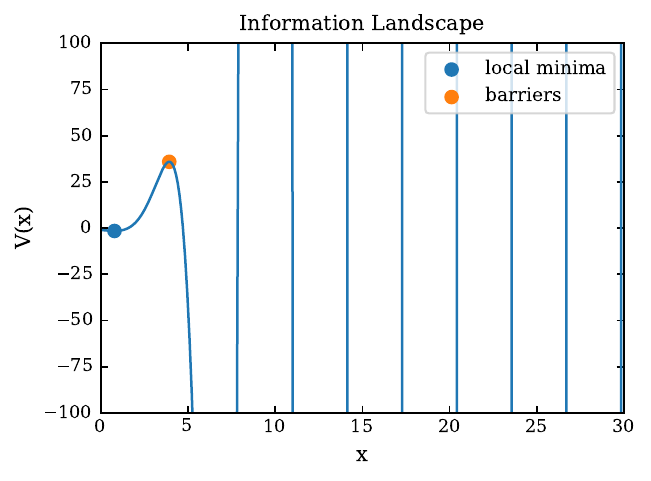}
    \caption{
    Information potential landscape for $\alpha=1$ over
    $0\leq x\leq30$, with the local extrema indicated. The increasing
    exponential envelope produces progressively deeper local minima and
    higher transition barriers as $x$ increases.
    }
    \label{fig:landscape_extrema}
\end{figure}

Figure~\ref{fig:landscape_extrema} shows that the third and subsequent extrema rapidly move beyond the scale of the first few extrema. Their potential values can exceed $100$ in magnitude within the plotted interval. This behaviour follows directly from the exponential envelope,

\begin{equation}
|V(x)|\sim e^{\alpha x}.
\end{equation}

Within the interpretation of the model, this increasing amplitude represents \textbf{increasing difficulty in accessing information states located farther from the origin}. Thus, the barrier between successive information states becomes increasingly difficult to cross as the information representation approaches the asymptotic intended-information state.

To examine the landscape structure without the extremely rapid growth produced by $\alpha=1$, a smaller value of the accessibility deterioration rate was also considered.

\begin{figure}[htbp]
    \centering
    \includegraphics[width=0.6\linewidth]{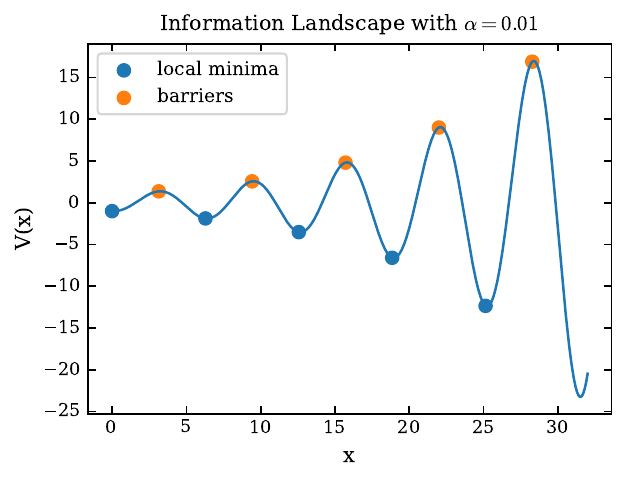}
    \caption{
    Information potential landscape for $\alpha=0.01$. The reduced
    accessibility deterioration rate produces a more slowly increasing
    exponential envelope, allowing the local minima and transition
    barriers to remain visually distinguishable over a larger range of
    the information coordinate.
    }
    \label{fig:landscape_alpha001}
\end{figure}

Figure~\ref{fig:landscape_alpha001} demonstrates that $\alpha$ controls the rate at which the landscape amplitude increases. A smaller $\alpha$ produces a less rapidly increasing barrier structure, while a larger $\alpha$ makes the potential landscape increasingly steep at larger $x$.

The effect of $\alpha$ is further demonstrated by comparing several values of the parameter.

\begin{figure}[htbp]
    \centering
    \includegraphics[width=\linewidth]{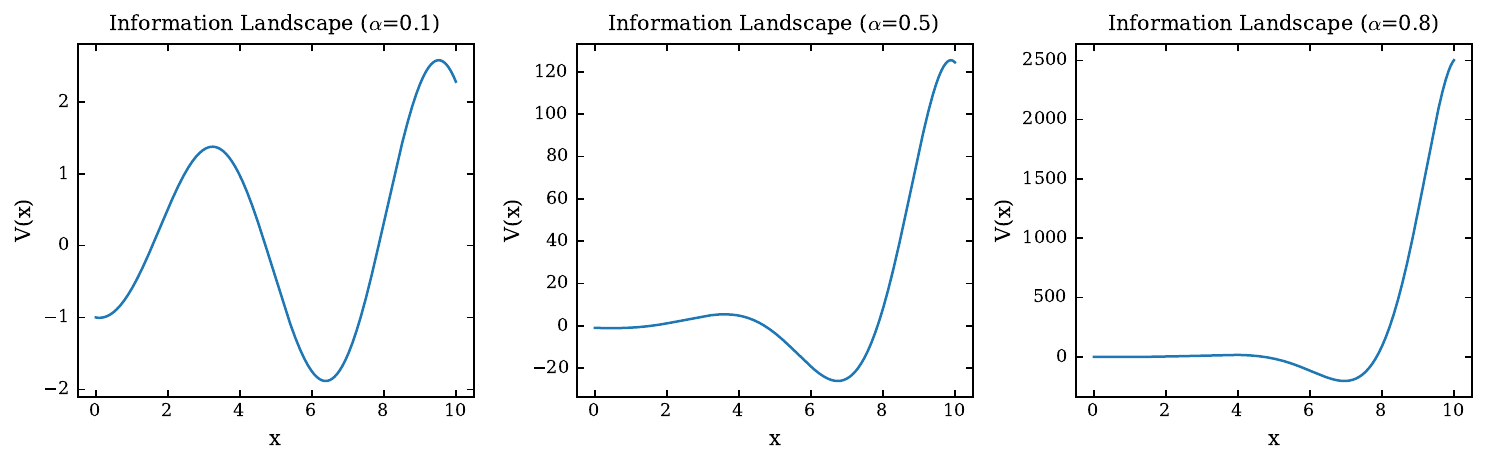}
    \caption{
    Comparison of information potential landscapes for different values
    of the accessibility deterioration rate $\alpha$. Increasing $\alpha$
    increases the exponential envelope and therefore increases the
    magnitude of both local minima and transition barriers at larger
    information coordinates.
    }
    \label{fig:landscape_comparison}
\end{figure}

As shown in Fig.~\ref{fig:landscape_comparison}, $\alpha$ strongly affects the potential magnitude at a given value of $x$. Importantly, $\alpha$ does not change the qualitative oscillatory structure of the landscape. Instead, it controls how rapidly the accessibility barrier grows along the information coordinate.

This provides a simple interpretation of $\alpha$ within the proposed model: it represents the rate at which information accessibility deteriorates as the representation moves farther from the origin and closer to the asymptotic intended-information state.

\subsection{Information Dynamics}

After defining the static information landscape, the temporal evolution of the information coordinate was investigated using Eq.~(\ref{eq:dynamics}). Three simplified information-processing regimes were considered: no update, constant update, and time-dependent update.

\begin{figure}[htbp]
    \centering
    \includegraphics[width=0.7\linewidth]{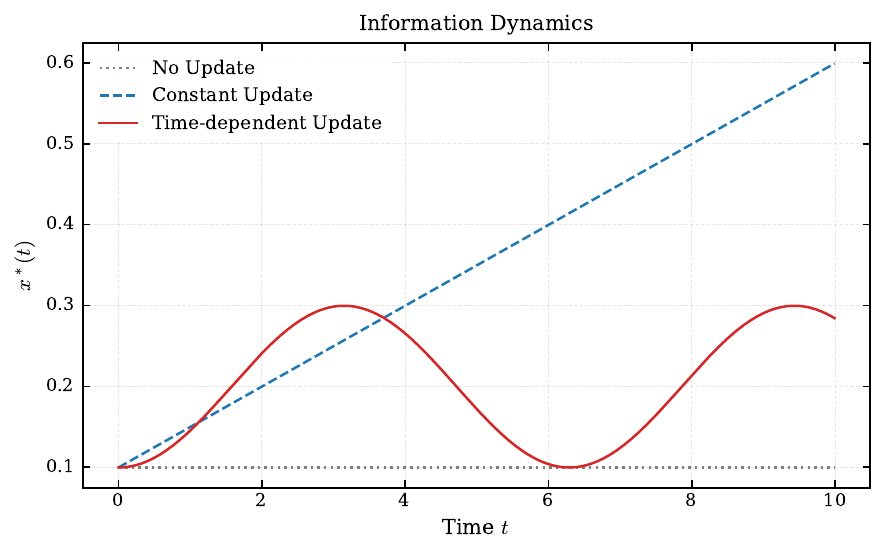}
    \caption{
    Simulated information trajectories for three information-processing regimes. The no-update case remains at its initial information state, the constant-update case changes its information coordinate at a constant rate, and the time-dependent case produces a non-monotonic trajectory due to the changing information-processing forcing.}
    \label{fig:information_dynamics}
\end{figure}

For the no-update case eq. \eqref{eq:noup},

\begin{equation}
x(t)=x_0.
\end{equation}
The information state remains unchanged because no information-processing forcing is applied. For the constant-update case Eq. \eqref{eq:constant}, the information coordinate therefore evolves as

\begin{equation}
x(t)=x_0+At.
\label{eq: constant_up}
\end{equation}

The resulting trajectory is linear, indicating that the information state moves through the information space at a constant rate. For the time-dependent case in Eq. \eqref{eq:timedep}, the information coordinate can move in both directions since it was modeled by a sinusoidal function. This represents a situation in which repeated processing of the same information topic can either improve or reduce the current information representation depending on the instantaneous state of the thought-processing function.

The simulation therefore illustrates the distinction between the static information landscape and information dynamics. The potential landscape defines the structure of possible stable information states, while $\Phi(t)$ determines how the information coordinate changes with time.

\subsection{Information Trajectory on the Potential Landscape}

\begin{figure}[htbp]
    \centering
    \includegraphics[width=0.6\linewidth]{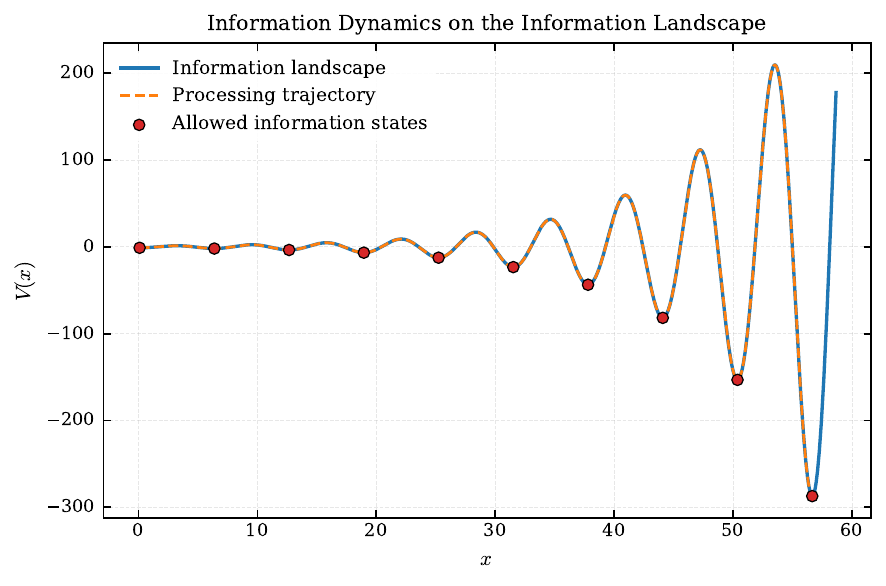}
    \caption{
    Information trajectory mapped onto the potential landscape for the simulated information dynamics. The landscape provides the static structure of stable information states and transition barriers, whereas the dynamic trajectory describes the temporal movement of the information representation through this landscape.}
    \label{fig:landscape_trajectory}
\end{figure}

The dynamic trajectories can also be mapped onto the potential landscape to visualize the movement of information through the proposed information space. For the constant-update case, the time required to reach a particular coordinate follows directly from Eq. \eqref{eq: constant_up} giving

\begin{equation}
t=\frac{x(t)-x_0}{A}.
\label{eq:constant_time}
\end{equation}
The corresponding trajectory is shown in Fig.~\ref{fig:landscape_trajectory}.

The figure illustrates that the trajectory and the landscape have different roles in the model. The landscape itself does not generate the information update. Instead, it defines the possible stable positions and the barriers separating them. The temporal movement is determined by the information-processing forcing $\Phi(t)$.

\subsection{Relationship Dynamics}

To investigate the evolution of the coupled systems, the discrepancy between their information states is evaluated after each information update. The discrepancy is used to describe how far the information representation of one system is from the information representation of the other system during the communication process. The resulting quantity is then used to compare different information-processing regimes and to investigate how changes in information separation can influence the subsequent relationship dynamics.

\subsubsection{Information Separation}

The discrepancy is normalized to facilitate comparison between the different information-update regimes. A normalized value of $0$ corresponds to the minimum separation observed within the selected simulation range, while a normalized value of $1$ corresponds to the maximum separation observed within the same range.

\begin{figure}[htbp]
    \centering
    \includegraphics[width=0.8\linewidth]{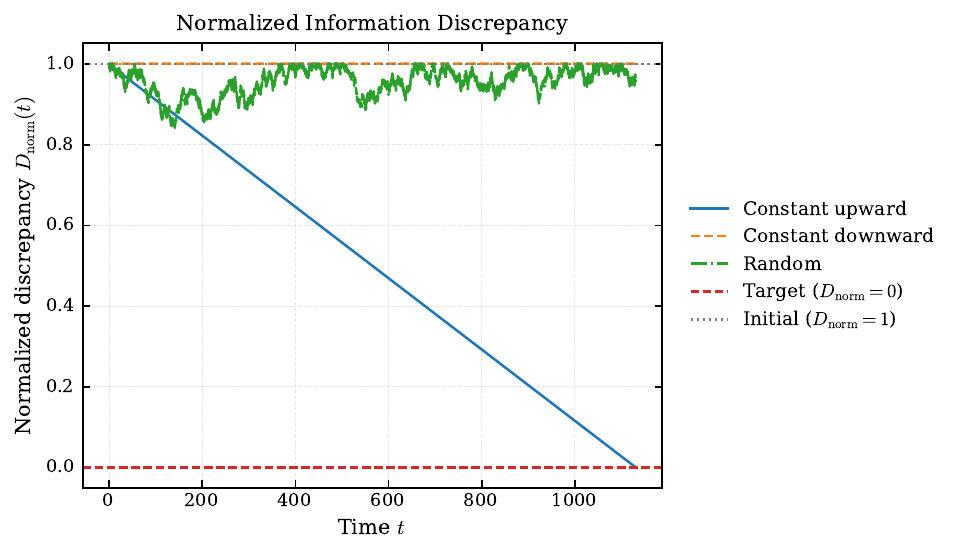}
    \caption{
    Normalized information discrepancy between the two coupled systems under constant upward, constant downward, and time-dependent random information processing. The discrepancy is normalized within the simulated range. The constant upward case approaches zero as the observed information reaches the intended-information state, while the constant downward case approaches and remains at the maximum separation allowed by the information-space boundary. The random case exhibits time-dependent fluctuations in the information separation.
    }
    \label{fig:normalized_discrepancy}
\end{figure}

The resulting trajectories in Fig. \ref{fig:normalized_discrepancy} shows three qualitatively different information-processing behaviours. In the constant upward case, the information discrepancy decreases continuously toward zero. This corresponds to the observed information moving toward the intended information state, such that the two information representations gradually become aligned. Once the observed information reaches the intended-information position in the numerical approximation, the discrepancy remains at zero since it could not exceed below it.

In contrast, the constant downward case produces an increasing separation until the observed information reaches the lower boundary of the information space. Since the intended information remains at a finite position representing the asymptotic limit and the observed information also constrained to $x\geq0$, the separation then remains at its maximum value within the simulated range. This case therefore represents a persistent mismatch between the intended and observed information states.

The time-dependent random case produces a qualitatively different behaviour. Instead of moving monotonically toward or away from the intended information, the information position fluctuates in time, resulting in a fluctuating information discrepancy. The random case therefore represents a genuinely time-dependent information process rather than a small perturbation around a single deterministic trajectory.

\subsubsection{Statistical Relationship Dynamics}

The information discrepancy can subsequently influence the relationship coupling, and the resulting relationship dynamics is evaluated for the five relationship states introduced in the model. For the statistical simulation, $N_{\mathrm{real}}=500$ independent realizations are generated. In each realization, the information-processing trajectory and the individual sentiment states are allowed to fluctuate in time. The resulting ensemble is then used to obtain the mean relationship dynamics and its standard deviation at each time.

\begin{figure}[htbp]
    \centering
    \includegraphics[width=0.8\linewidth]{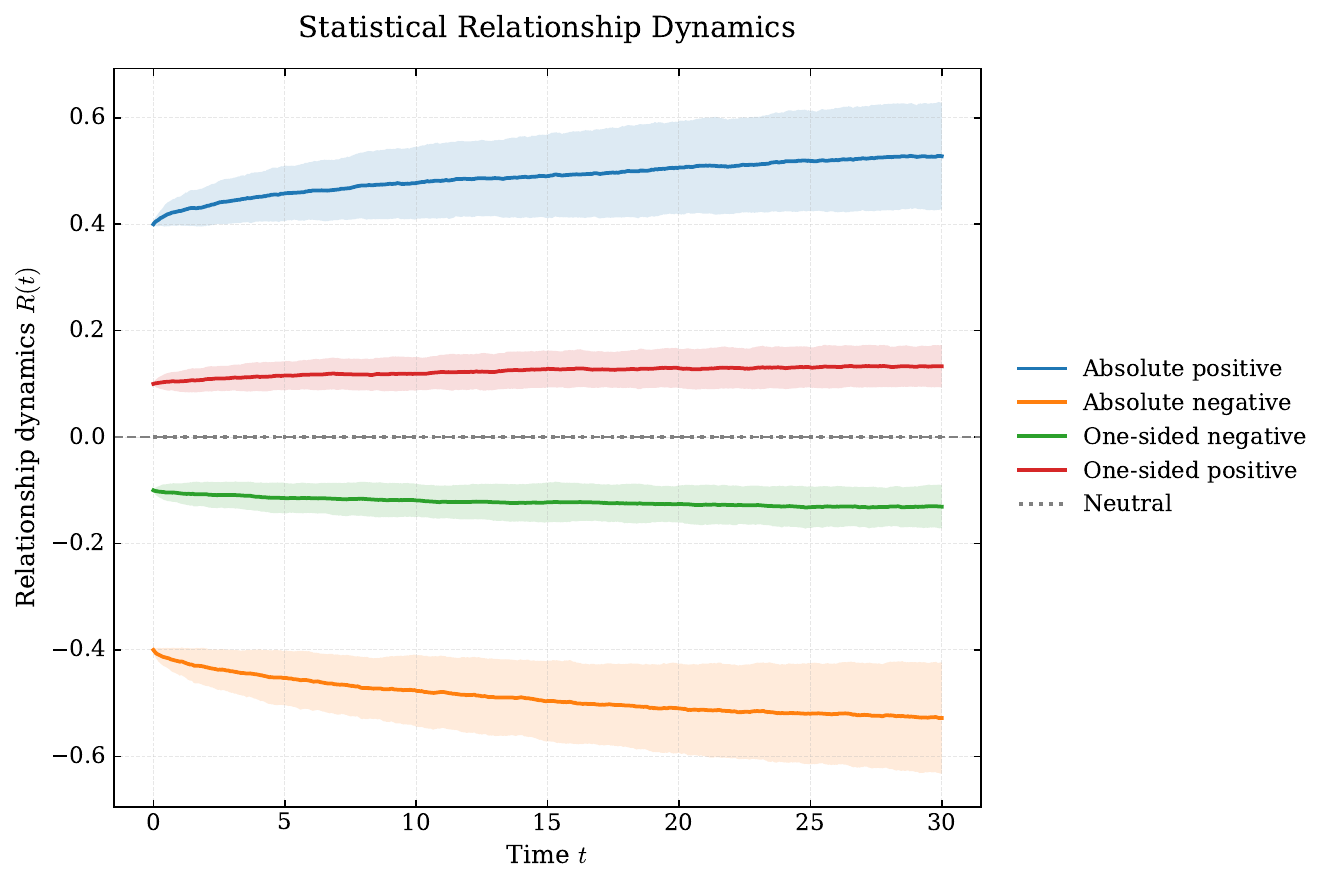}
    \caption{
    Statistical relationship dynamics for the five relationship states under fluctuating information processing. The solid curves represent the mean relationship dynamics over $N_{\mathrm{real}}=500$ realizations, while the shaded regions represent the corresponding $\pm1\sigma$ statistical fluctuations. The horizontal dashed line indicates the neutral state, $R(t)=0$. The positive and negative relationship regimes remain on their respective sides of the neutral state despite temporal fluctuations.
    }
    \label{fig:statistical_dynamics}
\end{figure}

The simulations verify that the proposed algebraic relationship preserves the five prescribed sentiment regimes under fluctuating information processing. Figure~\ref{fig:statistical_dynamics} shows the resulting average relationship dynamics for the five relationship states. The absolute positive and absolute negative cases exhibit the largest magnitudes, with the former remaining in the positive region and the latter remaining in the negative region throughout the simulation. This behaviour follows directly from the combined individual sentiment states, which remain strongly positive or strongly negative, respectively. The effect of the fluctuating information processing appears primarily through changes in the magnitude of the relationship dynamics rather than through a change in its sign.

The one-sided cases exhibit smaller magnitudes because the two individual sentiments partially compensate for each other. In the one-sided negative case, the negative sentiment is sufficiently strong that the combined sentiment remains negative, resulting in a relationship state below the neutral line. In the one-sided positive case, the positive sentiment dominates, resulting in a positive relationship state. Although both cases fluctuate in time, their mean trajectories remain separated from the neutral state. These one-sided cases could actually be interpreted as showing that relationship dynamics totally depend on the combination of both individuals' sentiments, along with the joint sentiment. When one party has a negative sentiment and the other has a positive sentiment, it doesn't necessarily mean that the relationship is deteriorated; however, it also strongly depends on how each of the sentiments affects each other.

The neutral case remains at $R(t)=0$ throughout the simulation. For this neutral control setup, $K_{12}$ is explicitly set to zero rather than evaluated from Eq. \eqref{eq:coupling}, allowing the uncoupled limit to be examined independently of the information-separation dynamics. This case is distinct from the other states because no effective relational dynamics is generated, even though individual internal sentiment states could, in principle, exist. Within the proposed model, this corresponds to a scenario where zero effective coupling prevents individual sentiments from translating into observable relationship dynamics effectively representing a lack of information sharing, with no observed state joining the information landscape. Consequently, regardless of the underlying sentiments, the relationship remains neutral in the total absence of communication.

In an indirect communication, however, such as telling gossip to other people or sharing another public story, the relationship could be no longer neutral since there is some information shared with indirect observers \cite{Shaw2011GossipNetworks, Sommerfeld2007GossipIndirectReciprocity, Pfeiffer2012ReputationGossip, Dores2021GossipReputationEverydayLife}. Another "indirectly" observed information could join the landscape, and therefore affect the value of the relationship's effective coupling strength $K_{12}$.

The widening statistical bands with time show that the magnitude of the relationship dynamics becomes increasingly sensitive to the accumulated fluctuations in the information-processing and sentiment states. However, the five regimes remain qualitatively distinguishable, indicating that the sign and approximate magnitude of the combined sentiment continue to determine the region in which the relationship dynamics evolves. The statistical result therefore demonstrates that the proposed relationship dynamics can preserve distinct relational regimes while allowing temporal fluctuations around their respective mean trajectories.

\begin{figure}[htbp]
    \centering
    \includegraphics[width=0.75\linewidth]{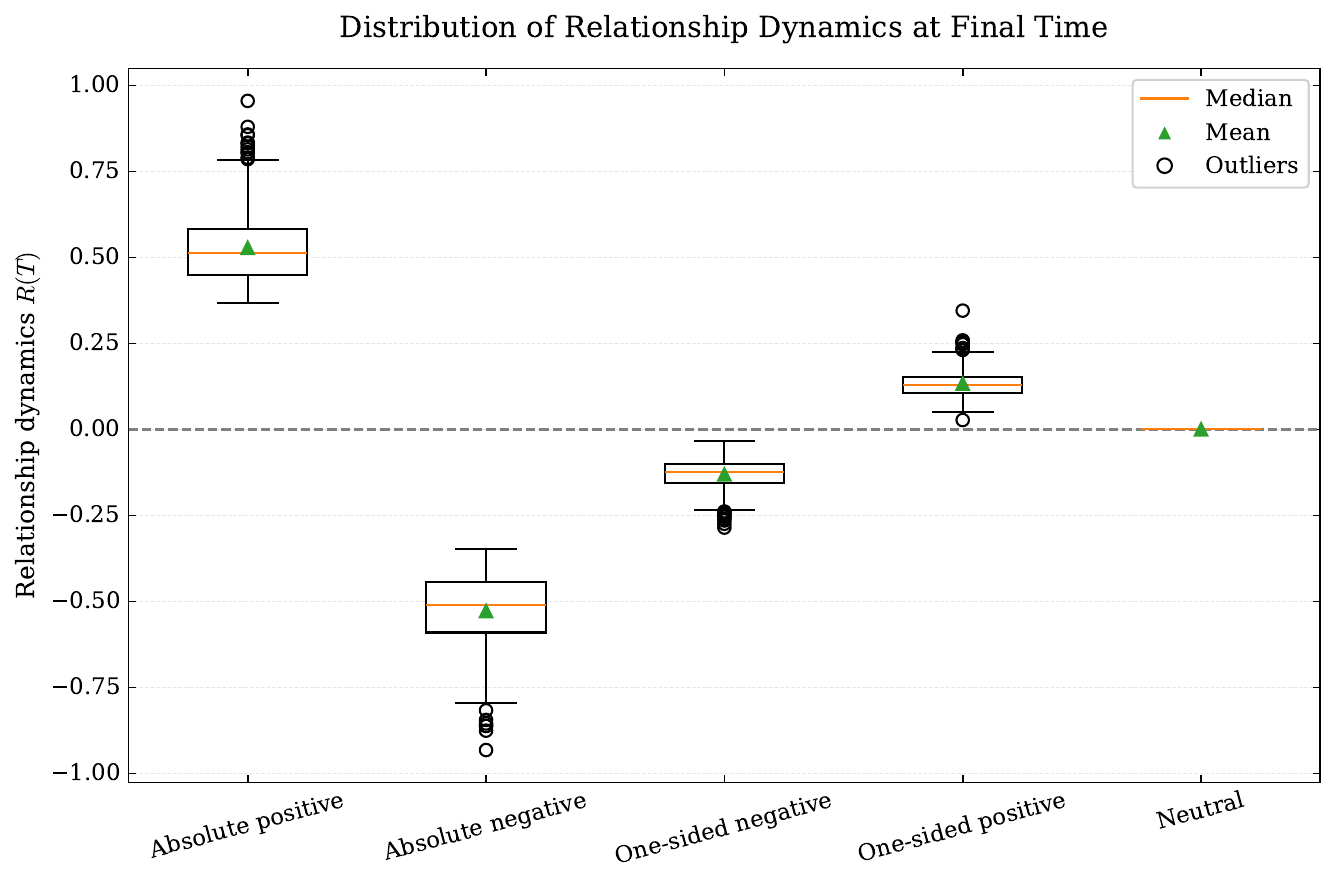}
    \caption{
    Distribution of the relationship dynamics at the final simulation time for the five relationship states. The box indicates the interquartile range, the horizontal line denotes the median, the triangular marker denotes the mean, and the circles indicate statistical outliers. The distributions remain centered in the expected positive, negative, or neutral regions according to the corresponding relationship state.
    }
    \label{fig:distribution_final_time}
\end{figure}

The distribution of the relationship dynamics at the final simulation time is shown in Fig.~\ref{fig:distribution_final_time}. The distributions are clearly separated according to the sign and magnitude of the combined individual sentiment. The absolute positive and absolute negative cases show the largest positive and negative central values, respectively, while the one-sided cases are located closer to the neutral state. The one-sided negative case remains below zero, whereas the one-sided positive case remains above zero. The neutral case collapses to a single value at zero, as expected from the absence of effective coupling.

The spread of each distribution reflects the fluctuations generated during the repeated information-processing process. In particular, the presence of outliers indicates that individual realizations can temporarily deviate substantially from the average relational state. Nevertheless, the central tendency of each distribution remains consistent with the qualitative relationship regime assigned to it. This provides an additional statistical view of the temporal results shown in Fig.~\ref{fig:statistical_dynamics}.

\subsubsection{An Illustrative Scenario}

\begin{figure}[htbp]
    \centering
    \includegraphics[width=0.8\linewidth]{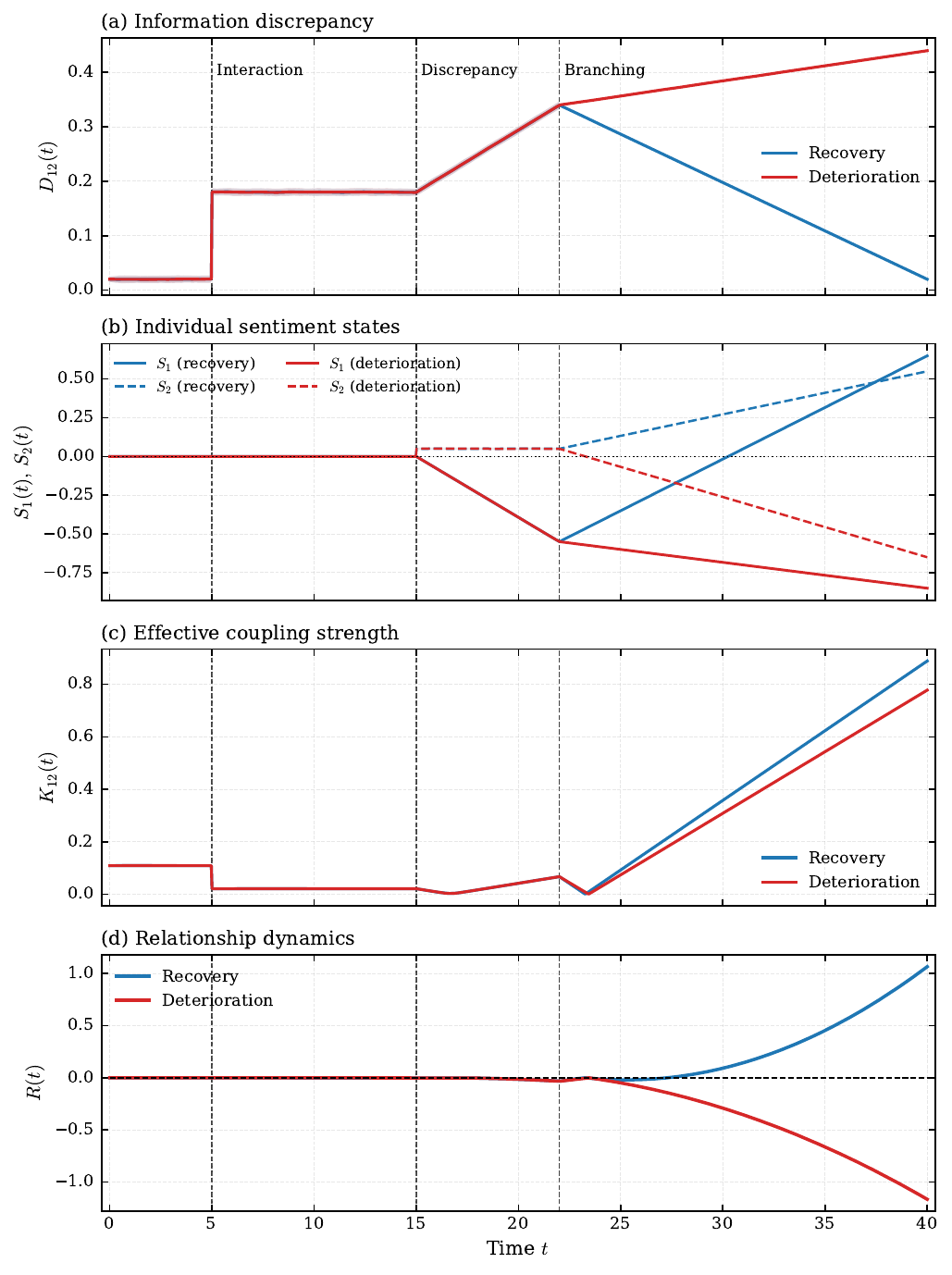}
    \caption{
    Illustrative evolution of a coupled relationship from an initially near-neutral state into two qualitative branches. The panels show (a) information discrepancy, (b) individual sentiment states, (c) effective coupling strength, and (d) relationship dynamics. Vertical dashed lines indicate the onset of interaction, the emergence of information discrepancy, and the subsequent branching point. Solid curves represent the recovery branch, while the corresponding curves for deterioration are shown for comparison. The trajectory illustrates how changes in information separation, baseline coupling, and individual sentiment can produce qualitatively different relationship outcomes from a similar initial state.
    }
    \label{fig:relationship_dynamics_scenario_statistical}
\end{figure}

To understand how the relationship dynamics evolves as a complete process, a more illustrative scenario is presented in Fig.~\ref{fig:relationship_dynamics_scenario_statistical}. The simulation begins with two systems that have not yet established an effective interaction. The resulting relationship dynamics therefore remains close to the neutral state, as the effective coupling is initially very small. At the beginning of the interaction, $t=5$, small fluctuations appear in the information discrepancy, but the relationship dynamics remains close to zero because the individual sentiment states are still approximately neutral.

At $t=15$, a noticeable information discrepancy begins to emerge. This indicates that the observed information is no longer remaining close to the previously aligned information state. At the same time, the sentiment of one system begins to move toward the negative region. The information discrepancy continues to increase until the branching point at $t=22$, where two qualitatively different relationship trajectories are considered. 

The final-state values for both branches are summarized in Table~\ref{tab:final_state_values}. The recovery branch reaches a state with a small information discrepancy and a relatively strong effective coupling, while both individual sentiments become positive. As a result, the relationship state reaches a positive value of $R=1.067\pm0.004$. In contrast, the deterioration branch ends with a larger information discrepancy and a lower effective coupling, while both individual sentiments become negative, resulting in a negative relationship state of $R=-0.987\pm0.003$.
\begin{table}[htbp]
    \centering
    \caption{
    Final-state values of the information discrepancy $D_{IO}$, effective
    coupling strength $K_{12}$, individual sentiments $S_1$ and $S_2$, and
    relationship state $R$ for the recovery and deterioration branches.
    Values are given as the mean $\pm$ standard deviation over the
    realizations at the final simulation time.
    }
    \label{tab:final_state_values}
    \begin{tabular}{lcc}
        \hline
        Quantity & Recovery & Deterioration \\
        \hline
        $D_{IO}$ & $0.020 \pm 0.002$ & $0.440 \pm 0.002$ \\
        $K_{12}$ & $0.889 \pm 0.001$ & $0.658 \pm 0.001$ \\
        $S_1$    & $0.650 \pm 0.003$ & $-0.850 \pm 0.003$ \\
        $S_2$    & $0.550 \pm 0.003$ & $-0.650 \pm 0.003$ \\
        $R$      & $1.067 \pm 0.004$ & $-0.987 \pm 0.003$ \\
        \hline
    \end{tabular}
\end{table}

A small bump in the effective coupling strength $K_{12}(t)$ can be observed around $t \gtrsim 22$. Since this function is constrained to always be positive, when the branching starts, which is when both individuals start to respond to the information discrepancy, its value initially decreases toward zero. The relationship therefore becomes temporarily neutral at an instant in time before the recovery and deterioration branches increase the coupling strength and make the relationship state more distinct.

In the recovery branch, the information discrepancy begins to decrease after the branching point. The two information representations therefore become increasingly aligned again. At the same time, the baseline coupling $K_0$ increases, representing the accumulation or strengthening of positive relational factors such as familiarity, shared experience, or other factors that help maintain the coupling. The initially negative sentiment also gradually recovers and becomes positive. These simultaneous changes increase the effective coupling and shift the combined sentiment toward the positive region. Consequently, the relationship dynamics gradually turns upward and eventually develops a strongly positive state.

In the deterioration branch, the information discrepancy continues to increase instead of decreasing. The baseline coupling $K_0$ is kept approximately constant, so no additional positive reinforcement of the relationship coupling is introduced. At the same time, both individual sentiment states progressively become negative. The combined sentiment therefore remains negative, while the remaining effective coupling continues to transmit this negative state into the relationship dynamics. The resulting trajectory moves progressively farther into the negative region.

The scenario therefore illustrates that the emergence of an information discrepancy does not uniquely determine the final relationship state. Instead, the subsequent evolution depends on how the information discrepancy, coupling strength, and individual sentiment states evolve together. A similar intermediate state can therefore lead to either recovery or deterioration depending on the subsequent dynamical response of the coupled systems. The constructed trajectories here, however, are not intended to represent the full complexity of real interpersonal relationships. Rather, they provide an illustrative realization of how the assumptions of the present model can lead to qualitatively different relationship outcomes.

The relationship state is therefore determined by the combined effect of coupling and individual sentiment rather than by the magnitude of the coupling alone. The end result of the relationship state, however, really depends on how both individuals manage the information discrepancy. If they can maintain the miscommunication without developing a negative sentiment, the relationship could remain stable. Vice versa, the opposite could also occur. Therefore, the influence of the thought-process is really important in determining how the information moves through the information landscape. Having a stable mood, good emotional control, and experience in decision-making \cite{Gross1998EmotionRegulation, Bloch2014EmotionRegulationMarital, Mazzuca2019EmotionRegulationMarital, CameronOverall2018EmotionRegulationVariability, Kohn2014NeuralModelCognitiveER} may influence the evolution of interpersonal sentiment and therefore represents a potentially important extension of the present model. Such an extension could further investigate under what conditions relationship deterioration may progress toward relationship break-up.

\section{Conclusion}

A physics-inspired interpersonal communication model is proposed based on several main assumptions: \textbf{intended and observed information}, in which the intended information is hardly accessible and is therefore represented as an asymptotic limit, while the observed information is already transformed through environmental transmission effects and the thought-processing function of the receiving system; \textbf{information potential landscape}, where information can remain in metastable states surrounded by potential barriers in an abstract information space; and \textbf{information transmission}, which describes the dynamics of information as it is transmitted and processed by different systems. The state of the \textbf{relationship dynamics} between two systems is then defined as the combined effect of the coupling strength between the systems and the individual sentiment of each system. Numerical implementation of the model demonstrates the information potential landscape, information dynamics, and information trajectories using a finite approximation, and also provides an illustrative relationship scenario in which the relationship can either recover or deteriorate after a conflict emerges.

\section{Limitations and Further Works}

The present model is intentionally a minimal toy model and has several limitations.

\begin{enumerate}

    \item \textbf{The information coordinate is reduced to one dimension}. Human information and interpretation are inherently multidimensional, and a one-dimensional coordinate cannot represent the full structure of semantic, emotional, contextual, and relational information. Further work could extend the information coordinate into a multidimensional space.

    \item \textbf{The thought-processing function $F_i$ is treated as a black box}. Consequently, the current model does not identify which specific cognitive or psychological mechanisms produce a particular information trajectory. Further work could investigate the explicit form of this function and the mechanisms through which a system may access information beyond the directly observed state. For example, individuals in closer relationships may sometimes understand each other more effectively without explicit verbal communication. Possible mechanisms for describing such transitions could include quantum-like tunnelling or the expansion into additional dimensions, allowing information to cross a potential barrier without fully traversing its classical potential height.

    \item \textbf{The potential function $V(x)$ is phenomenological and physics-inspired} rather than empirically derived from measurements of human communication. The exponential envelope and oscillatory structure are chosen to represent increasing barriers and multiple stable information states. In further work, the potential landscape could take a different functional form, while the concept of stable information states could remain.

    \item \textbf{The current simulations use simplified forcing functions}. The constant and random cases are mathematical test cases rather than realistic models of human cognition. Further work could implement more detailed stochastic or nonlinear dynamical models, including possible applications of chaos theory.

    \item \textbf{The relationship dynamics is currently demonstrated using a single scenario with assigned initial values}. Further work could introduce empirically motivated initial conditions and evaluate the relationship dynamics using real communication data. For example, sentiment analysis and machine-learning methods could be used to estimate the individual sentiment states and evaluate the proposed relationship dynamics.
    
\end{enumerate}







\section*{Declaration of Competing Interest}
The author declares that there are no known competing financial interests or personal relationships that could have appeared to influence the work reported in this paper.

\section*{Acknowledgements}

The author would like to thank Prof. Rinto Anugraha NQZ for a brief discussion about this work.

\printcredits

\bibliographystyle{model1-num-names}

\bibliography{cas-refs}



\end{document}